\documentclass[pdflatex, sn-basic, Numbered]{sn-jnl}

\usepackage{graphicx}%
\usepackage{multirow}%
\usepackage{amsmath,amssymb,amsfonts}%
\usepackage{amsthm}%
\usepackage{mathrsfs}%
\usepackage[title]{appendix}%
\usepackage{xcolor}%
\usepackage{textcomp}%
\usepackage{manyfoot}%
\usepackage{booktabs}%
\usepackage{algorithm}%
\usepackage{algorithmicx}%
\usepackage{algpseudocode}%
\usepackage{listings}%
\usepackage{comment}

\theoremstyle{thmstyleone}%
\theoremstyle{thmstyletwo}%

\theoremstyle{thmstylethree}%

\begin{document}

\title[Article Title]{Automatic brain tumor segmentation in 2D intra-operative ultrasound images using  magnetic resonance imaging tumor annotations}


\author*[1,2]{\fnm{Mathilde Gajda} \sur{Faanes}}\email{mathilde.faanes@sintef.no}

\author[1,3]{\fnm{Ragnhild Holden} \sur{Helland}}

\author[4,5]{\fnm{Ole} \sur{Solheim}}

\author[1]{\fnm{S\'ebastien} \sur{Muller}}

\author[1,3]{\fnm{Ingerid} \sur{Reinertsen}}

\affil*[1]{\orgdiv{Department of Health Research}, \orgname{SINTEF Digital},\city{Trondheim}, \country{Norway}}

\affil[2]{\orgdiv{Department of Physics}, \orgname{Norwegian University of Science and Technology (NTNU)}, \city{Trondheim}, \country{Norway}}

\affil[3]{\orgdiv{Department of Circulation and Medical Imaging}, \orgname{Norwegian University of Science and Technology (NTNU)}, \city{Trondheim}, \country{Norway}}

\affil[4]{\orgdiv{Department of Neurosurgery}, \orgname{St. Olavs hospital}, \city{Trondheim}, \country{Norway}}

\affil[5]{\orgdiv{Department of Neuromedicine and Movement Science}, \orgname{Norwegian University of Science and Technology (NTNU)}, \city{Trondheim}, \country{Norway}}


\abstract{\noindent 
Automatic segmentation of brain tumors in intra-operative ultrasound (iUS) images could facilitate localization of tumor tissue during resection surgery. The lack of large annotated datasets limits the current models performances. In this paper, we investigated the use of tumor annotations in magnetic resonance imaging (MRI) scans, which are more accessible than annotations in iUS images, for training of deep learning models for iUS brain tumor segmentation. We used 180 annotated MRI scans with corresponding unannotated iUS images, and 29 annotated iUS images. Image registration was performed to transfer the MRI annotations to the corresponding iUS images before training the nnU-Net model with different configurations of the data and label origins. The results showed no significant difference in Dice score for a model trained with only MRI annotated tumors compared to models trained with only iUS annotations and both, and to expert annotations, indicating that MRI tumor annotations can be used as a substitute for iUS tumor annotations to train a deep learning model for automatic brain tumor segmentation in iUS images. The best model obtained an average Dice score of $0.62\pm0.31$, compared to $0.67\pm0.25$ for an expert neurosurgeon, where the performance on larger tumors were similar, but lower for the models on smaller tumors. In addition, the results showed that removing smaller tumors from the training sets improved the results. The main models are available here: \url{https://github.com/mathildefaanes/us_brain_tumor_segmentation/tree/main}
}




\maketitle

\section{Introduction}
\label{sec:introduction}
Diffuse gliomas are primary brain tumors that infiltrate normal brain tissue, making complete surgical resection impossible \cite{delgado-lopez_diffuse_2017}. Although incurable, treatment can prolong life and improve quality of life and brain functions \cite{kheirollahi_brain_2015}. Surgical resection is often the preferred primary treatment option, as the extent of tumor resection is linked to prolonged survival \cite{sanai_glioma_2008}. However, removing or damaging healthy surrounding tissue can negatively impact patient function \cite{sastry_applications_2017}. Precise tumor border localization is therefore crucial for successful resection.

Low-cost, real-time intra-operative ultrasound (iUS) has been shown to enhance the surgical outcome \cite{cepeda_non-navigated_2024}. However, the images have a limited field of view and often contain noise and artifacts that make them difficult to interpret \cite{carton_image_2021}. Automatic brain tumor segmentation in iUS images could facilitate interpretation and help the surgeon perform a more complete resection.

Deep learning can address this segmentation task as demonstrated in the CuRIOUS-SEG challenge in 2022 \cite{xiao_lesion_2023}. The winning contribution came from Qayyum et al. \cite{qayyum_segmentation_2023} who achieved a Dice score of 0.57 using 23 annotated iUS images from RESECT \cite{xiao_retrospective_2017} and RESECT-SEG \cite{behboodi_resect-seg_2022} as training set, which are to date, the only publicly available annotated dataset. Recently, Dorent et al. \cite{dorent_miccai} introduced a patient-specific segmentation model trained on simulated ultrasound images from pre-operative MRI scans from a single patient. Their approach obtained a median Dice score of 0.84-0.87, compared to 58.5-71.7 for a general model trained on the RESECT dataset \cite{dorent_miccai}. Despite good results, their approach is limited to pre-resection scenarios and requires time-consuming and computationally extensive training for each patient. Having a model that generalizes well to all patients would be more advantageous in the clinic, but the lack of large annotated iUS datasets limits further development.

To address this issue, we explored whether tumor annotations in pre-operative MRI scans could be used as a substitute for manual tumor delineations in iUS images to expand the training set of a deep learning model for automatic brain tumor segmentation in iUS images. MRI tumor annotations are more accessible than iUS tumor annotations, either from published datasets \cite{juvekar_remind_2023}, or by software for automatic tumor segmentation such as Raidionics \cite{bouget_raidionics_2023}. Since non-navigated 2D ultrasound imaging is more commonly used and low-cost than 3D ultrasound imaging \cite{cepeda_non-navigated_2024}, this work focuses on 2D images. Training sets with only MRI tumor annotations, only iUS tumor annotations and both MRI and iUS tumor annotation were used to train 2D deep learning models. All models were evaluated on a separate test set with manual iUS annotations, and compared to the inter-observer variability.

\section{Methods}
\begin{figure}[!ht]
    \centering
    \includegraphics[width=0.85\columnwidth]{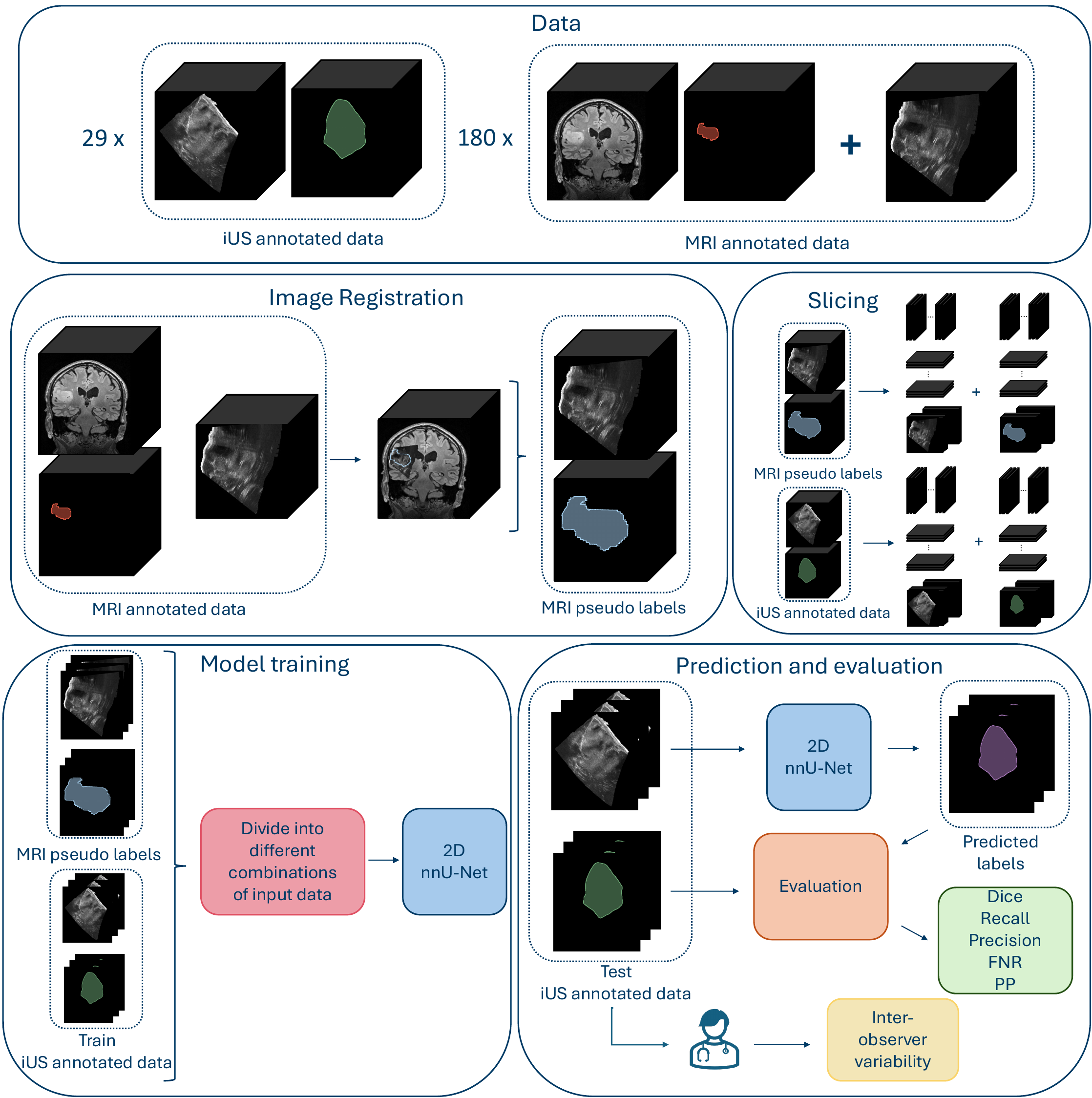}
    \caption{Overview over the data, data preparation, model training and evaluation of the results.}
    \label{fig:oversikts_figur}
\end{figure}

An overview of the methods is shown in Fig. \ref{fig:oversikts_figur}.

\subsection{Data}\label{sec2}
Both annotated and un-annotated iUS images were used in this study. The data were divided into two categories: iUS annotated data and MRI annotated data, as explained in Fig. \ref{fig:oversikts_figur}. The iUS annotated data consisted of iUS images with manual tumor annotations, whereas the MRI annotated data comprised iUS images without manual tumor annotations, but with pre-operative MRI scans with tumor annotations. Depending on the experiment, all or parts of the data were used for training and testing.

The iUS annotated data were obtained from 29 patients with gliomas who underwent surgery at St. Olavs Hospital, Trondheim, Norway. Among these, 23 have been published as part of the REtroSpective Evaluation of Cerebral Tumors (RESECT) dataset \cite{xiao_retrospective_2017}, where the annotations are available from the RESECT-SEG dataset \cite{behboodi_resect-seg_2022}. The remaining 6 were used as test set in the CuRIOUS-SEG challenge, organized in conjunction with the MICCAI 2022 conference \cite{curious_2022_brain_2022}. The datasets contains iUS images acquired before, during and after resection. Only the images acquired before resection were used in this study, giving a total of 29 annotated 3D ultrasound images. 

The MRI annotated data consisted of 180 images from the Brain Resection Multimodal Imaging Database (ReMIND) \cite{juvekar_remind_2023}, and in-house data from St. Olavs University Hospital. From the ReMIND database, 103 iUS images acquired before resection with corresponding annotated pre-operative MRI scans from 55 patients with glioma or metastasis who underwent surgery for the first time, were included in the study. 

The in-house dataset from St. Olavs Hospital contained 77 pre-resection 3D iUS images with corresponding annotated pre-operative MRI scans from 43 patients with a glioma or metastasis who underwent surgery for the first time. The ultrasound images were mainly acquired with a 12FLA-L linear probe, using the Sonowand Invite neuronavigation system \cite{gronningsaeter_sonowand_2000} or the CustusX neuronavigation system \cite{askeland_custusx_2016}. The data was collected through the Central Norway Brain Tumor Registry and Biobank, or through several research projects on ultrasound-guided neurosurgery, with written informed consent from all patients \cite{cancer_registry_of_norway_norwegian_2023}. For cases with missing tumor annotations, the open-source software for automatic brain tumor segmentation in MRI scans, Raidionics \cite{bouget_raidionics_2023}, was used to generate annotations. This resulted in 29 annotated 3D iUS images and 180 un-annotated 3D iUS images with corresponding pre-operative 3D MRI scans and annotations from 127 patients.

\subsection{MRI-iUS registration}
To use the MRI tumor annotations as labels for the ultrasound images lacking annotations, the MRI tumor annotations were transferred to the corresponding ultrasound space by rigid registration using the medical image analysis software, ImFusion Suite (Version 2.42.2) \cite{imfusion_imfusion_2018}. All registrations were visually inspected to ensure adequate alignment.

\subsection{Model pre-processing, training and post-processing}
The nnU-Net framework was used in this study because it is a standardized and self-configuring deep learning framework, where optimal pre-processing, hyperparameters and post-processing are determined based on the given dataset. In addition, it has shown state-of-the-art performance in many biomedical image segmentation tasks \cite{isensee_nnu-net_2021}. 

The iUS 3D volumes were sliced in all three perpendicular directions and saved in the NIfTI-format to obtain 2D images. Only tumor-containing slices were kept. The 2D configuration of nnU-Net was used with the nnUNetTrainerDA5 option. All models were trained with five-fold cross-validation with early stopping with a patience of 30 epochs, ensuring no information leaks by keeping all 2D slices from one patient in the same fold. The pre-processing, hyperparameters, and post-processing proposed by the framework for each dataset, were used.

An Intel Core Processor (Broadwell, no TSX, IBRS) CPU with 16 cores, 64GB of RAM, Tesla V100S (32GB) dedicated GPU and a regular hard-drive was used for training. The nnU-Net framework v.2.2 was used with Python 3.8 and PyTorch v.2.2.0.

\subsection{Experiments}
\subsubsection{Comparison of tumor area cut-off values}
In this experiment, we used the iUS images with registered MRI labels from the MRI annotated data as explained in Fig. \ref{fig:oversikts_figur}. The 2D slices contained tumors with a vast variability in tumor areas. Slices from the edges of the tumor volumes were particularly small, and there was a higher risk of mismatch between the MRI pseudo label and the tumor in the iUS due to inaccurate image registration, compared to slices in the middle of the tumor volume. This experiment therefore studied different tumor area cut-off values, to investigate the effect of excluding small and possibly poorly aligned tumor labels that could be a source of noise rather than a valuable contribution to the training data. Nine models were trained in this experiment using 2D slices from the 180 3D MRI annotated images with tumor area cut-off values ranging from 0 to 300~mm$^2$. The number of 2D slices used for training can be seen in Table \ref{tab:MRI_tumor_area_test_set}. All models were evaluated on all tumor-containing slices of the 29 3D volumes of the iUS annotated data resulting in 14 107 test slices.

\subsubsection{Comparison between MRI labels, iUS labels and manual annotations}

In this experiment, the best model from the first experiment was compared to a model trained on only iUS annotated data, and a model trained on both iUS and MRI annotated data. For the iUS-model, the iUS annotated data was split into a training and test set. The 23 patients from RESECT were used for training, and the 6 patients from the test set of the CuRIOUS-SEG challenge were used as a test set, resulting in 2259 test 2D slices. For the MRI+US\_200 model, 8 3D images from the MRI annotated data were excluded from the training set due to poor image quality. The number of 2D slices used for training is shown in Table \ref{tab:MRI_tumor_area_test_set}. To ensure a fair comparison between the models, both models were trained using nnU-Net with a tumor area cut-off of 200 mm$^2$ based on the results from the first experiment. The models are available here: \url{https://github.com/mathildefaanes/us_brain_tumor_segmentation/tree/main}. 

In addition, the performances of the models were compared to the inter-observer variability, to assess the difficulty of the segmentation task. The first author manually annotated 3D images of the test set, using 3D Slicer (Version 5.2.2) \cite{fedorov_3d_2012}. The annotations were adjusted and validated by an experienced neurosurgeon (OS) and compared against the CuRIOUS-SEG ground truth annotations. These results, are in the following, referred to as "Annotator".

\subsection{Evaluation and statistical analysis}
To evaluate the models in each experiment, the pixel-wise segmentation performances were evaluated on the test sets by calculating Dice score, precision, recall and false negative rate (FNR) between the ground truth masks and the binary prediction masks. The percentage of positive predictions (PP), thus if one or more pixels were classified as a tumor in an image, was also calculated image-wise.

Linear regression models were used to assess statistical differences in Dice scores between the models in each experiment. The substantial within-patient correlation between 2D slices was taken into account by clustering on a patient level (Python package \textit{pyfixest}). To compensate for multiple pairwise comparisons between groups, the Bonferroni correction of type I error was applied. The effect size of the differences was given by Cohen's $d = (\bar{x_i} - \bar{x_j}) / s$, where $\bar{x_i}$ and $\bar{x_j}$ represent the average Dice score achieved with model $i$ and $j$, and $s$ is the pooled standard deviation calculated using clustered standard errors $S_i$ estimated by the regression \cite{cohen}. As all groups had the same number of observations $n$, it reduces to $s^2 = n (S_i^2 + S_j^2) / 2$. We used $d \sim 0.01$ describing very small and $d \sim 0.2$ describing small effect sizes \cite{cohen-sizes}.

In addition, the influence of the tumor area cut-off value in Experiment 1 was assessed using linear regression to predict Dice score, precision, recall and false negative rate (FNR), correcting for within-patient correlation using the same method. A similar logistic regression was used for the binary positive prediction (PP).

\section{Results}

\subsection{Comparison of tumor area cut-off values}

\begin{figure}[!ht]
    \centering
    \includegraphics[width=0.95\columnwidth]{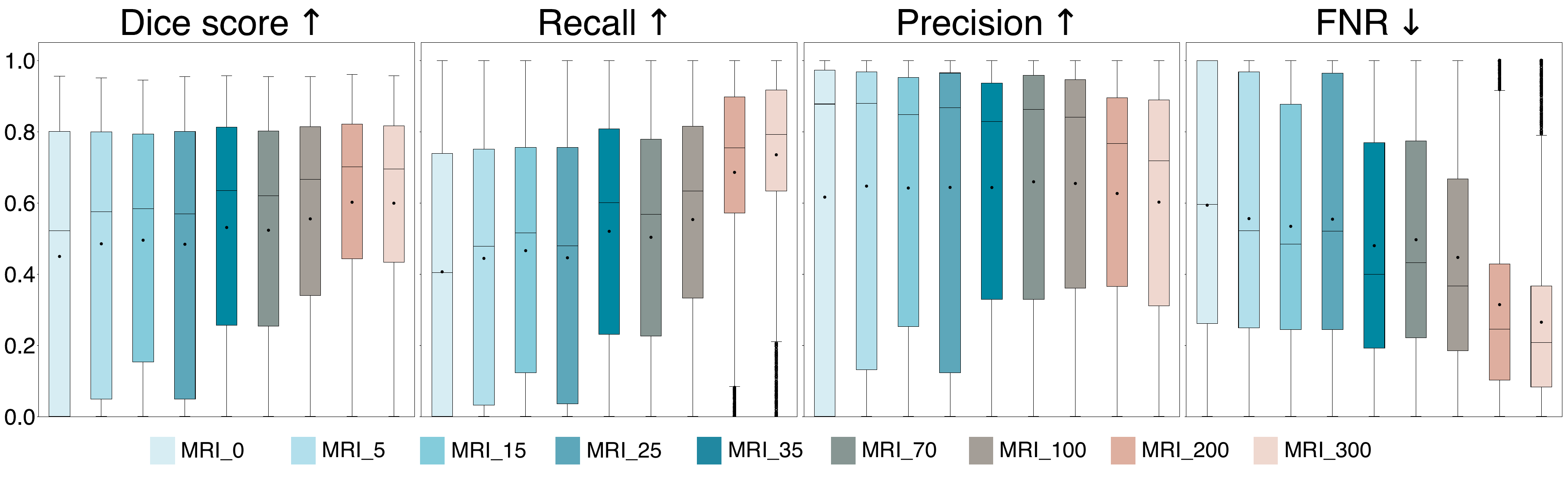}
    \caption{Comparison of Dice score, recall, precision and false negative rate (FNR) for the 9 models trained with different tumor area cut-off values, where MRI\_0 is trained using all tumor-containing slices and MRI\_5 to MRI\_300 are trained with slices containing a tumor area larger than 5 mm$^2$ up to 300 mm$^2$, respectively. Average values are represented by dots and median values are represented by lines. The arrows indicate which value is preferred.}
    \label{fig:Res-tumorarea-MRI}
\end{figure}

\begin{figure}[!ht]
    \centering
    \includegraphics[width=0.85\columnwidth]{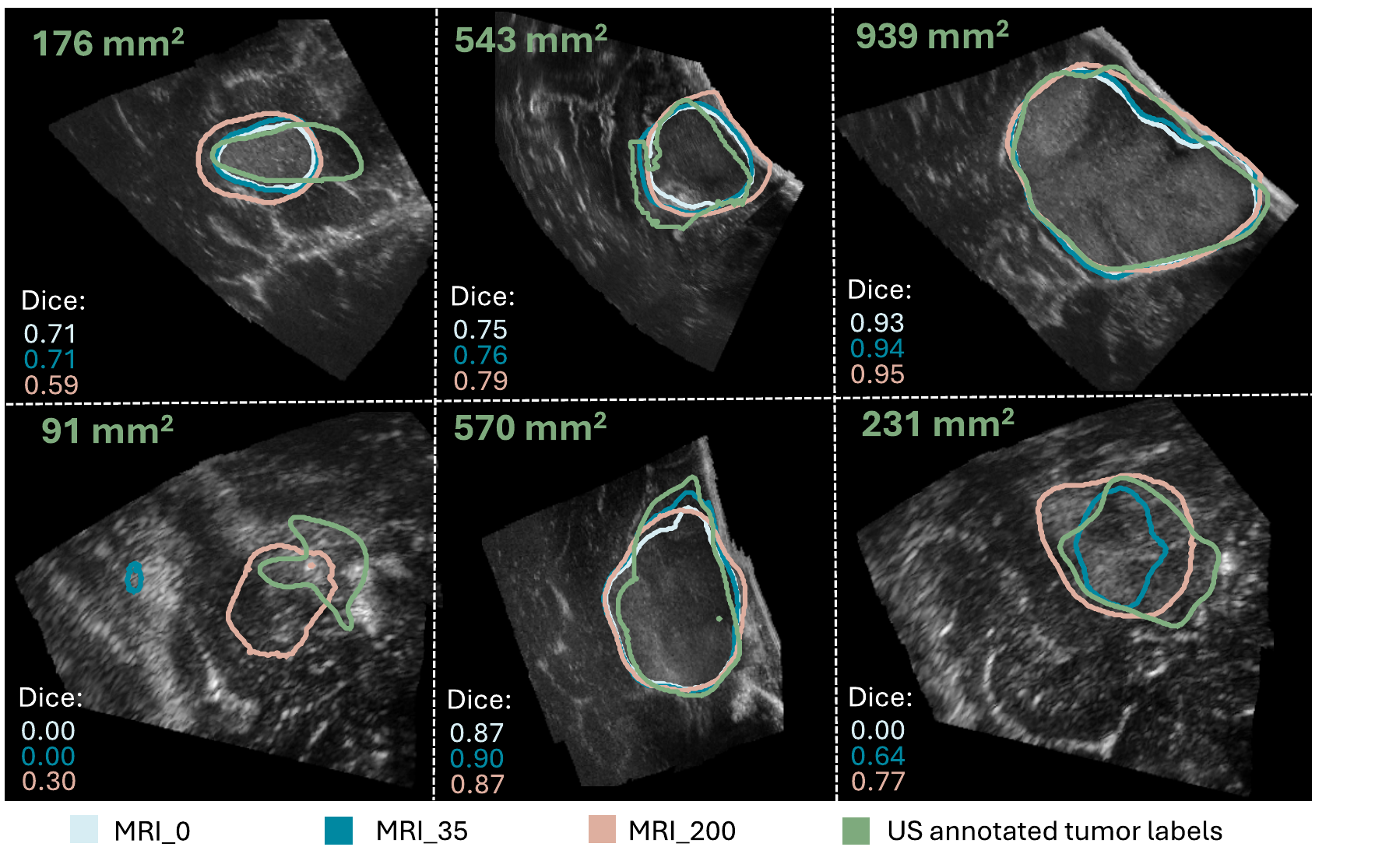}
    \caption{Comparison of the segmentation masks of the ground truth (in green), the MRI\_0 model (in light blue), the MRI\_35 model (in blue), and the MRI\_200 model (in beige). The ground truth tumor area and Dice scores are also shown for each case.}
    \label{fig:mri_ex}
\end{figure}

\begin{table*}[h!]
    \centering
    \caption{The Dice score (mean $\pm$ standard deviation) for the nine MRI annotated data models from Experiment 1 in the upper part of the table, and for the three models and the Annotator from Experiment 2, in the lower part, evaluated on the test 2D slices from each experiment in different tumor area intervals; between 0-35 mm$^2$, 35-200 mm$^2$ and above 200 mm$^2$, and on the total amount of test slices. For Experiment 1, $^*$ is showing statical significance from MRI\_0 ($P<0.0064$, Bonferroni) with effect sizes in parentheses. The percentage of positive predictions (PP) for the total test set and the number of training samples (N) written in thousands, are also shown. The arrows indicate which value is preferred. }
    \begin{tabular}{l|cccc|cc}
    \hline
    \multirow{3}{*}{} &
    \multicolumn{3}{c}{\textbf{Tumor area ranges [mm$^2$]}}&\multirow{3}{*}{}\\ 
    \textbf{Model} 
    &\textbf{0-35 $\uparrow$}
    &\textbf{35-200 $\uparrow$}
    &$\mathbf{> 200\ \uparrow}$
    &\textbf{Total $\uparrow$} 
    &\textbf{PP $\uparrow$} 
    &\textbf{N} \\
    \hline
    MRI\_0 &0.03$\pm$0.10&0.20$\pm$0.28&0.61$\pm$0.30&0.45$\pm$0.36 (-) &72.9 \% &105 \\
    MRI\_5&0.05$\pm$0.13&0.26$\pm$0.30&0.63$\pm$0.27&0.49$\pm$0.34$^*$ (0.07) &79.2 \% &100 \\
    MRI\_15&0.06$\pm$0.13&0.28$\pm$0.29&0.64$\pm$0.25&0.50$\pm$0.33$^*$ (0.10)&84.6 \%&95  \\
    MRI\_25&0.05$\pm$0.13&0.26$\pm$0.30&0.63$\pm$0.27&0.48$\pm$0.34$^*$(0.07) &78.9 \% &92 \\
    MRI\_35&0.07$\pm$0.15&0.31$\pm$0.30&0.68$\pm$0.23&0.53$\pm$0.32$^*$(0.18)&87.6 \% &89\\
    MRI\_70&0.06$\pm$0.13&0.32$\pm$0.30&0.67$\pm$0.23&0.52$\pm$0.32$^*$(0.16)&86.9 \%&80 \\
    MRI\_100&0.07$\pm$0.13&0.34$\pm$0.29&0.70$\pm$0.21&0.56$\pm$0.31$^*$(0.23)&90.0 \% &73\\
    MRI\_200& \textbf{0.07$\pm$0.09}& \textbf{0.40$\pm$0.25}&0.75$\pm$0.15& \textbf{0.60$\pm$0.28$^*$ (0.25)}&98.6 \% & 51\\ 
    MRI\_300&0.06$\pm$0.07&0.38$\pm$0.21& \textbf{0.76$\pm$0.13}& 0.60$\pm$0.27$^*$ (0.21) &\textbf{99.8} \% & 35\\
    \hline

    MRI\_200&0.06$\pm$0.09&0.33$\pm$0.27&0.79$\pm$0.15& 0.58$\pm$0.32 &96.6 \% &51\\
    MRI+US\_200&0.08$\pm$0.10&0.40$\pm$0.27&\textbf{0.81$\pm$0.12}&0.62$\pm$0.31 & 99.0 \% & 57\\
    US\_200&0.08$\pm$0.10&0.37$\pm$0.22&0.77$\pm$0.12&0.59$\pm$0.29&\textbf{100 \%} & 8\\ 
    Annotator&\textbf{0.20$\pm$0.23}&\textbf{0.61$\pm$0.26}&0.77$\pm$0.14&\textbf{0.67$\pm$0.25} &95.2 \% & - \\
    \hline
    \end{tabular}
    \label{tab:MRI_tumor_area_test_set}
\end{table*}

Fig. \ref{fig:Res-tumorarea-MRI} shows a comparison of the evaluation metrics Dice score, recall, precision, and FNR obtained on all 14 107 2D slices from the 29 iUS annotated patients for the nine models trained with different tumor area cut-off values, represented by box plots. Dice and recall increased significantly for increasing cut-off ($P<10^{-5}$), FNR decreased significantly ($P<10^{-9}$), whereas precision showed no significant trend ($P=0.071$). Fig. \ref{fig:mri_ex} shows examples of predicted segmentation masks from the MRI\_0, MRI\_35, and MRI\_200 model and the ground truth, represented with the same colors as in the box plots. The Dice scores and ground truth tumor areas are also shown for each case. 

The upper part of Table \ref{tab:MRI_tumor_area_test_set} shows the average Dice score and standard deviation for the entire and different tumor area ranges of the test set for the models. In addition, it shows the percentage of positive predictions in the entire test set and the number of training samples N for each model. Statistical significance and effect size of the models compared to the MRI\_0 model are also shown. The linear regression showed that the percentage of positive prediction (PP\%) increased significantly with increasing tumor area cut-off ($P<10^{-16}$).

\subsection{Comparison between MRI labels, iUS labels and inter-observer variability}
\begin{figure}[!ht]
    \centering
    \includegraphics[width=0.95\columnwidth]{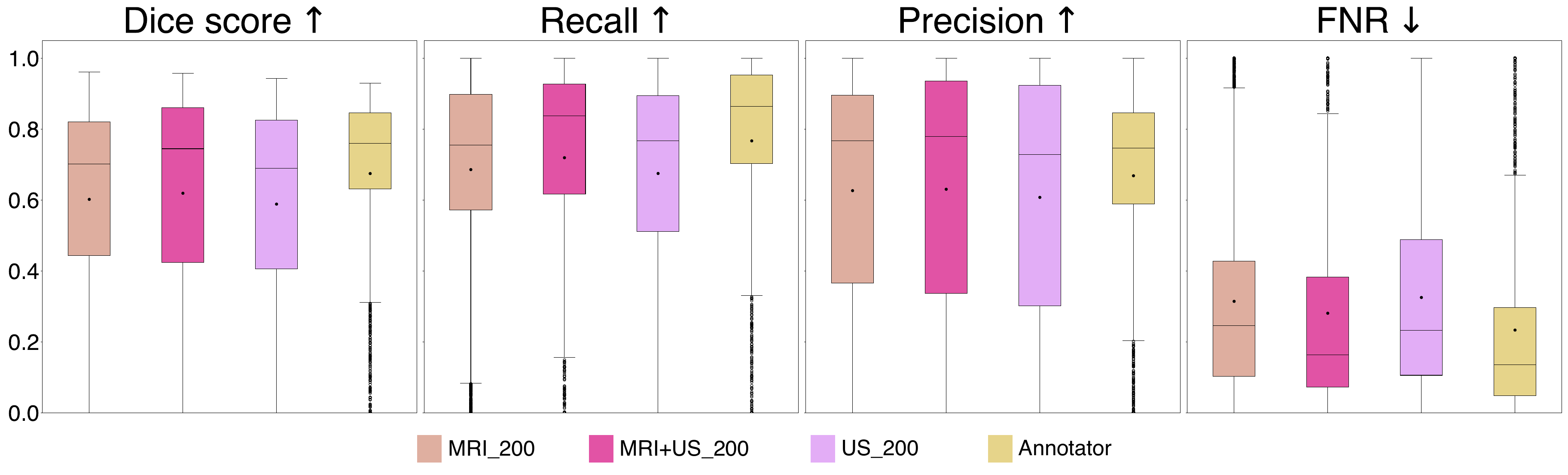}
    \caption{Comparison of Dice score, recall, precision and false negative rate (FNR) for the MRI\_200, MRI+US\_200, and US\_200 models, and for the Annotator. The arrows indicate which value is preferred.}
    \label{fig:comp-res}
\end{figure}

Fig. \ref{fig:comp-res} shows a comparison of the evaluation metrics Dice, recall, precision, and FNR, represented by box plots, for the MRI\_200 model, MRI+US\_200 model, US\_200 model, and the Annotator. 

\begin{figure}[!ht]
    \centering
    \includegraphics[width=0.85\columnwidth]{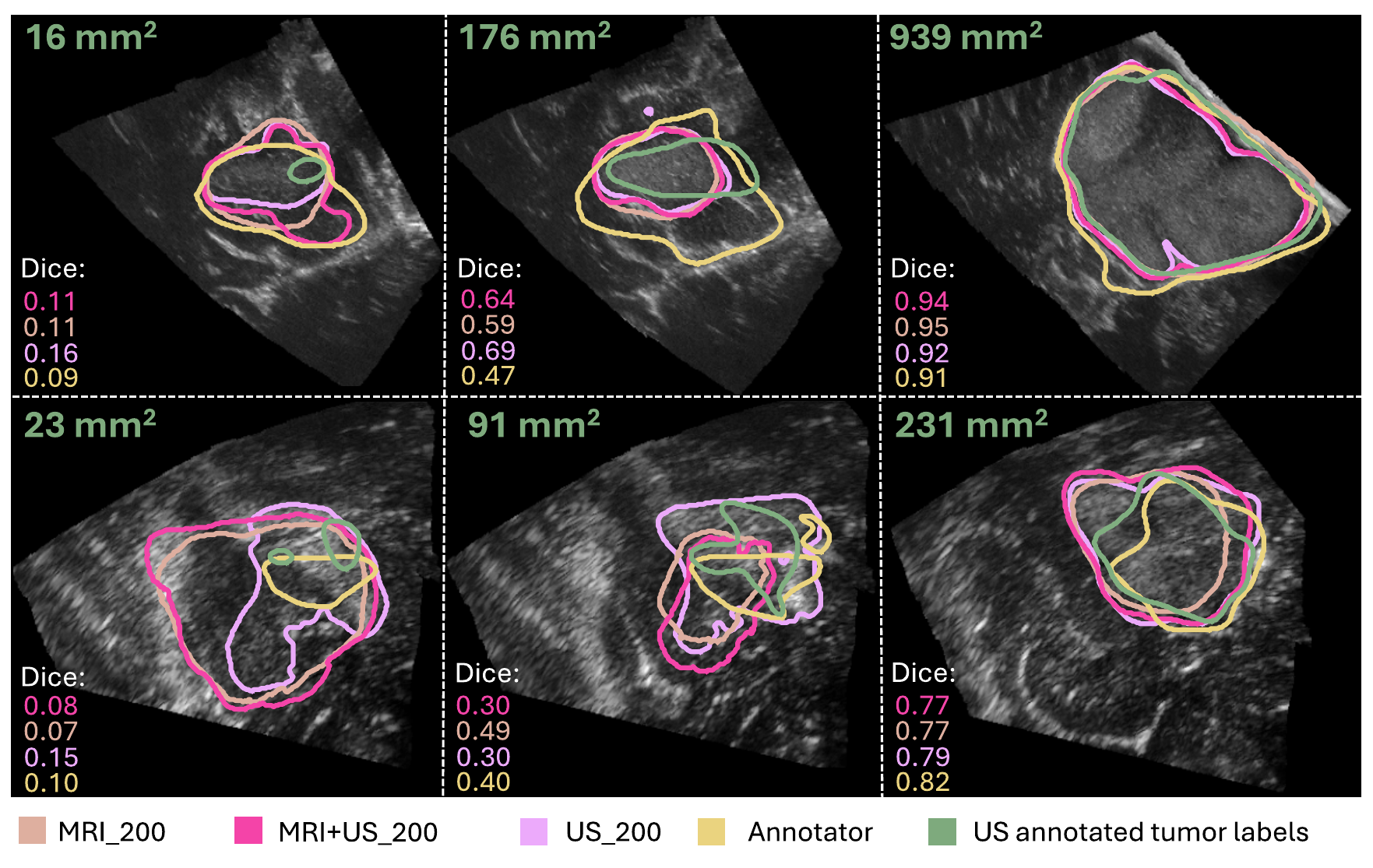}
    \caption{Comparison of the segmentation masks of the ground truth (in green), the Annotator (in yellow), the MRI\_200 model (in beige), the MRI+US\_200 model (in pink), and the US\_200 model (in purple). The ground truth tumor area and Dice scores are also shown for each case.}
    \label{fig:seg_exp}
\end{figure}

Fig. \ref{fig:seg_exp} shows examples of segmentation masks from the models and the Annotator, represented with the same colors as in the box plots, and the ground truth for test slices with different tumor areas.

The lower part of Table \ref{tab:MRI_tumor_area_test_set} shows the average Dice score and standard deviation for different tumor area ranges of the test set, and for the total test set, with statistical results shown in Table \ref{tab:cohen_P}, for the second experiment. In addition, Table \ref{tab:MRI_tumor_area_test_set} shows the percentage of positive predicted cases of the test set and the number of training samples (N) for each model. 

\begin{table}[h]
    \centering
    \caption{P-values from a pairwise comparison between the Dice scores on the test set for the MRI\_200, MRI+US\_200, US\_200 models and the Annotator are shown in the upper right part of the table, where P$>0.0085$ (Bonferroni) indicates statistical significance. The effect size of the differences are quantified by Cohen's d, with d$ \sim0.01$ indicating very small effect size and d$ \sim0.2$ indicating small effect sizes, are shown on the bottom left side.}
    \begin{tabular}{l|cccc}
    \hline
    & MRI\_200 & MRI+US\_200 & US\_200 & Annotator \\
    \hline
    MRI\_200 & - & 0.062 & 0.823 & 0.247 \\
    MRI+US\_200 & 0.036 & - & 0.085 & 0.420 \\
    US\_200 & 0.005 & 0.036 & - & 0.235 \\
    Annotator & 0.047 & 0.030 & 0.047 & - \\
    \hline
    \end{tabular}
    \label{tab:cohen_P}
\end{table}

\section{Discussion}
The purpose of this study was to investigate whether MRI tumor annotations can successfully be used as a replacement for iUS tumor annotations, for training an automatic brain tumor segmentation model for iUS images.

In the first experiment, we found that the models' performance improves by increasing tumor area cut-off values in the training set as evidenced by a significant increase in Dice score, recall and PP, and a significant decrease in FNR ($P<0.0064$, Bonferroni) compared to using all tumor areas (MRI\_0), shown in Fig. \ref{fig:Res-tumorarea-MRI} and Table \ref{tab:MRI_tumor_area_test_set}. Precision remains unchanged ($P>0.0064$, Bonferroni), suggesting that the ability to detect tumor pixels improves with larger tumor area cut-off values. This is illustrated in Fig. \ref{fig:mri_ex}, showing similar segmentation masks for all models except the MRI\_0 and MRI\_35 models where the tumor is missed in some cases. In addition, we found that a tumor area cut-off value around 200 mm$^2$ seems to provide the best and most stable results as indicated by that improvement in Dice score stagnates with a tumor cut-off value at 200  mm$^2$, regardless of tumor area in the test set, and by that the MRI\_200 model has the highest effect size, as shown in Fig. \ref{fig:Res-tumorarea-MRI} and Table \ref{tab:MRI_tumor_area_test_set}. This value was therefore used in the second experiment.

A possible explanation for the improvement in model performance when excluding the smallest tumors from the training set could be that the inaccuracies in the image registration have a larger impact on the smaller tumors, and including these might add more noise to the training data rather than valuable information. This illustrates that higher quality over quantity in the training data might be a good compromise in this case, as including the smaller tumors increases the training sample size, but decreases the performance. In addition, the segmentation task itself could be more challenging for smaller tumors because the slices with smaller tumor areas are often from the tumor border which is infiltrating healthy tissue, and could thus give a poorer contrast in the ultrasound images than the larger tumors areas. This can be seen in Fig. \ref{fig:mri_ex}. Additionally, segmentation of small structures is a well-known challenge in medical image segmentation because the amount of tumor pixels is low compared to background pixels, giving a high class-imbalance. Furthermore, Dice score, which is used in the loss function of nnU-Net combined with cross-entropy, is highly sensitive to errors in small structures \cite{isensee_nnu-net_2021}, and including small structures in the training could be disadvantageous for the network update.

From the results in the second experiment, we found indistinguishable performances from models trained with different label origins, evidenced by similar scores and no significant differences (P$>0.0085$, Bonferroni) and very small effect sizes (d$\sim0.01$), as shown in Table \ref{tab:MRI_tumor_area_test_set}, Fig. \ref{fig:comp-res} and Table \ref{tab:cohen_P}. The average Dice scores were $0.62\pm0.31$,  $0.59\pm0.28$, and $0.58\pm0.32$ for the MRI+US\_200, US\_200, and MRI\_200 model, respectively. This indicates that MRI tumor annotation can be used as labels for ultrasound images. As a results, clinicians can save valuable time by reducing time spent on manually annotating ultrasound images for developing deep learning models. However, the results did not improve even though the MRI+US\_200 model was trained on a dataset approximately seven times larger compared to the US\_200 models. Table \ref{tab:MRI_tumor_area_test_set} and Fig. \ref{fig:seg_exp} show that the models in this experiment also have poor results on small tumors, which could indicate that small tumors limit further improvements.

Regarding the comparison of the deep learning models' performance to the Annotator's performance, the Annotator achieved superior scores for all metrics on the entire test set, although no statistically significant differences were found (P$>0.0085$, Bonferroni), and the effect sizes were small (d$\sim0.01$). However, the statistical analysis was limited by the size of the test set. Although the number of 2D slices was large, the small patient sample (n=6) makes the performance highly sensitive to the test patients. A larger test set with diverse tumor types, annotated by multiple experts, would be in favor of a more robust statistical analysis, and better evaluation of the models. Nevertheless, the Dice scores obtained on the larger tumors ($>$ 200 mm$^2$) were similar to the scores achieved by the deep learning models, indicating an expert-level performance for larger tumors. However the segmentation of small structures remains a challenge, as illustrated in Fig. \ref{fig:seg_exp}.

Compared to previous work, Qayyum et al. \cite{qayyum_segmentation_2023} obtained an average Dice score of 0.57 on the same test set in 3D, compared to 0.62 for the MRI+US\_200 model in 2D. Even though this is not directly comparable, it indicates that our approach achieves comparable or better results, and supports the usage of MRI tumor annotations. The patient-specific approach of Dorent et al. \cite{dorent_miccai}, achieved better results with a median Dice score of 0.84-0.87, compared to a median Dice score of 0.74 for the MRI+US\_200 model. However, these models were fine tuned to each patient in the training dataset, requiring an amount of time and resources that makes a clinical adaption of the models challenging.

Despite promising results using MRI tumor annotations as a substitute for iUS annotations, the task of intra-operative brain tumor segmentation remains challenging. Although the models achieved promising performances on larger tumors, more thorough testing and validation is needed before clinical use. In addition, the detection of small structures remains a challenge. It is crucial for extending the model to segmentation of tumor tissue during and after resection, and should be the focus of future work. An overall limitation of this study is the small number of patients. Tumor annotations in both pre- and post-operative MRI scans from patients who have undergone brain tumor resection surgery, with corresponding iUS images, can be included to increase the number of smaller tumors in the training set. Different methods for improving the registration of the MRI labels to the iUS tumor, such as affine or non-linear registration methods or manual adjustment, could be explored. Additionally, loss functions tailored for small structures, a fundamentally different deep learning network like a vision transformer network, or foundation models like USFM \cite{jiao_usfm_2024}, could be tested. Other strategies than semantic segmentation could also be investigated, such as bounding boxes proposed by Weld et. al \cite{weld_challenges_2024}.

\section{Conclusion}

Although the detection of small tumors remains a challenge, our study showed that MRI tumor annotations can successfully be used as substitute labels for unannotated iUS images to train an automatic brain tumor segmentation model for 2D iUS images.

\section*{Acknowledgment}
\subsection*{Competing Interests} All the authors declare that they have no conflict of interest.

\subsection*{Ethical approval} This study has been approved and performed in accordance with ethical standards. 

\subsection*{Informed consent} Informed consent was obtained from all participants in this study.

\bibliography{biblio}

\end{document}